\documentclass[superscriptaddress,twocolumn]{revtex4-1}
\usepackage{amsmath}
\usepackage{graphicx}

\begin{document}

\title{Following Strain-Induced Mosaicity Changes of Ferroelectric Thin Films by Ultrafast Reciprocal Space Mapping}

\author{D. Schick}\email{daniel.schick@uni-potsdam.de}
\author{A. Bojahr}
\author{M. Herzog}
\affiliation{Institut f\"ur Physik \& Astronomie, Universit\"at Potsdam, Karl-Liebknecht-Str. 24-25, 14476 Potsdam, Germany}
\author{P. Gaal}
\affiliation{Helmholtz-Zentrum Berlin f\"ur Materialien und Energie GmbH, Wilhelm-Conrad-R\"ontgen Campus, BESSY II, Albert-Einstein-Str. 15, 12489 Berlin, Germany}
\author{I. Vrejoiu}
\affiliation{Max-Planck-Institut f\"ur Mikrostrukturphysik, Weinberg 2, 06120 Halle, Germany}
\author{M. Bargheer}
\affiliation{Institut f\"ur Physik \& Astronomie, Universit\"at Potsdam, Karl-Liebknecht-Str. 24-25, 14476 Potsdam, Germany}

\date{\today}

\begin{abstract}
We investigate coherent phonon propagation in a thin film of ferroelectric PbZr$_{0.2}$Ti$_{0.8}$O$_3$ (PZT) by ultrafast x-ray diffraction (UXRD) experiments, which are analyzed as time-resolved reciprocal space mapping (RSM) in order to observe the in- and out-of-plane structural dynamics simultaneously. 
The mosaic structure of the PZT leads to a coupling of the excited out-of-plane expansion to in-plane lattice dynamics on a picosecond timescale, which is not observed for out-of-plane compression.
\end{abstract}

\maketitle


Oxides are attractive constituents of future nano-electronic devices, because of their broad spectrum of outstanding physical properties, such as ferroelectricity and ferromagnetism, and owing to the progress made in the fabrication of high quality epitaxial heterostructures.\cite{schl2008a}
Epitaxial strain engineering and the careful choice of mechanical and electrical boundary conditions enable a direct influence on these  functionalities.\cite{han2011a,bass2007a,xu1991a,jia2011a,leek2001a}
Structural defects and nanoscale inhomogeneities, such as dislocations and domains, typically affect the properties of functional oxides and have been extensively studied by experiment and theory.\cite{dago2005a,fitz1991a}
Ultrafast x-ray diffraction (UXRD) emerged as a powerful tool to observe lattice motion in real time\cite{rous2001a,barg2006b,cher2009a} and has provided a deeper insight in the structure-property relations of functional oxides on ultrashort timescales.
Recent femtosecond x-ray scattering  experiments on ferroelectric oxides showed that electron screening induces an ultrafast piezoelectric response of the lattice\cite{dara2012a} and that in turn the deformation leads to a change of the polarization.\cite{korf2007b} 
However, these experiments were conducted on rather perfect epitaxial crystals.
The influence of nano-domains has been considered in experiments on transient phases\cite{ichi2011a}, but the role of static structural defects remained unexplored on such ultrafast timescale.

Here we exemplify how ultrafast reciprocal space mapping (URSM) using a laser-based plasma x-ray source yields direct additional information on the reversible in-plane structure dynamics in a ferroelectric perovskite PbZr$_{0.2}$Ti$_{0.8}$O$_3$ (PZT) film which is solely induced by the existence of dislocations typical of such materials.
In particular, the width of the PZT Bragg reflection reports that tensile out of plane strain leads to drastically increased damping. The energy flows into in-plane strain which is evidenced by the in-plane component of the reciprocal space map.
Our results indicate that in mismatched epitaxial films of oxide materials, with their high susceptibility to the formation of domains and dislocations, in-plane phenomena emerge on a hundred picosecond timescale.
URSM yields the relevant information on lateral lattice dynamics in such materials in which nanoscale inhomogeneities inherently broaden the peaks in reciprocal space.
It is important to realize that such inhomogeneities are a natural paradigm in oxides originating from competing phases with similar free energy rather than a result of imperfect crystal growth.\cite{dago2005a}
A better understanding of such time-domain effects in novel functional oxide materials will be important to study the influence of structural defects on the ultrafast response of collective phenomena, such as piezoelectricity.




\begin{figure}
\includegraphics[width=0.9\columnwidth]{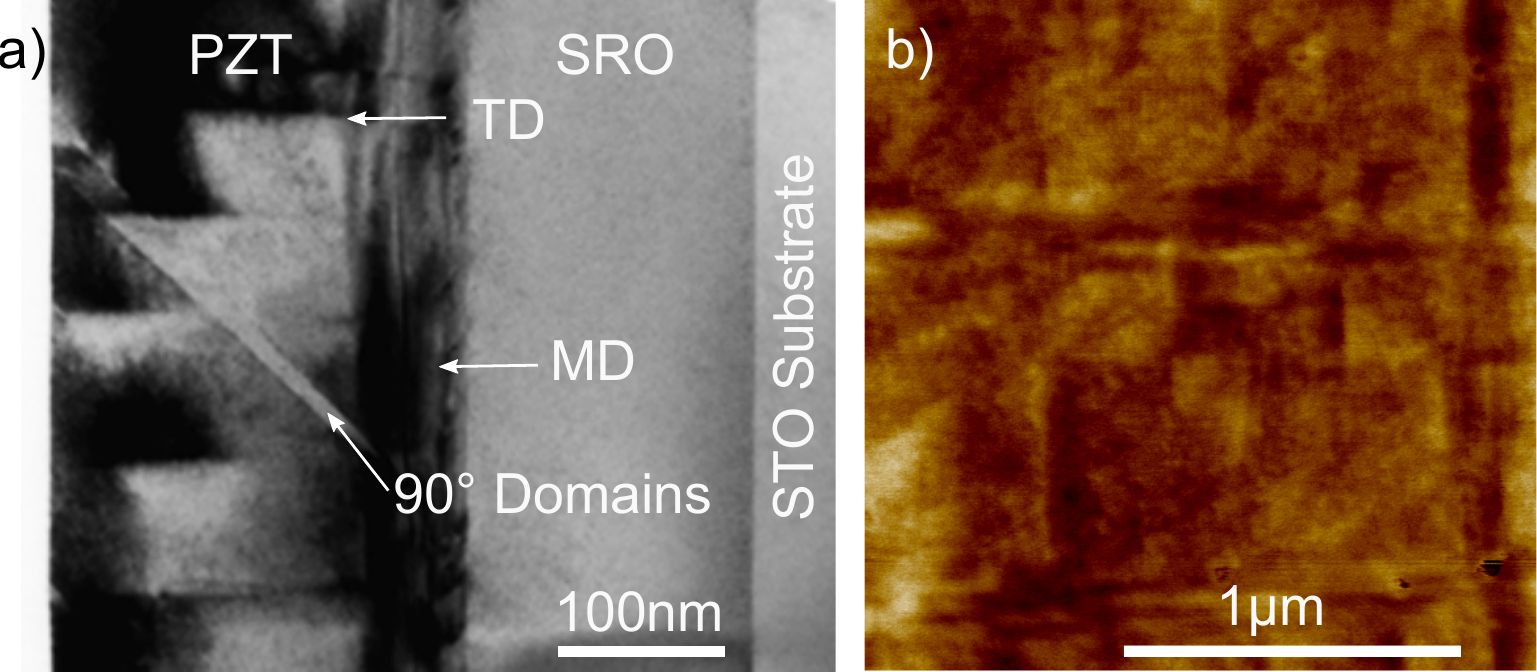}
\caption{TEM and AFM images of the PZT-SRO double layer grown onto an STO substrate by PLD. 
a) The cross section TEM micrograph reveals that a minority of $90^{\circ}$ $a$-domains are embedded in the matrix of the $c$-axis grown tetragonal PZT film, as proved by the AFM topography image in b) as well. 
Misfit dislocations (MD) and threading dislocations (TD) formed at the SRO-PZT interface and across the PZT film, accounting for the lateral inhomogeneity on a sub-100~nm length scale.
}
\label{fig:temafm}
\end{figure}

\begin{figure}
\includegraphics[width=0.9\columnwidth]{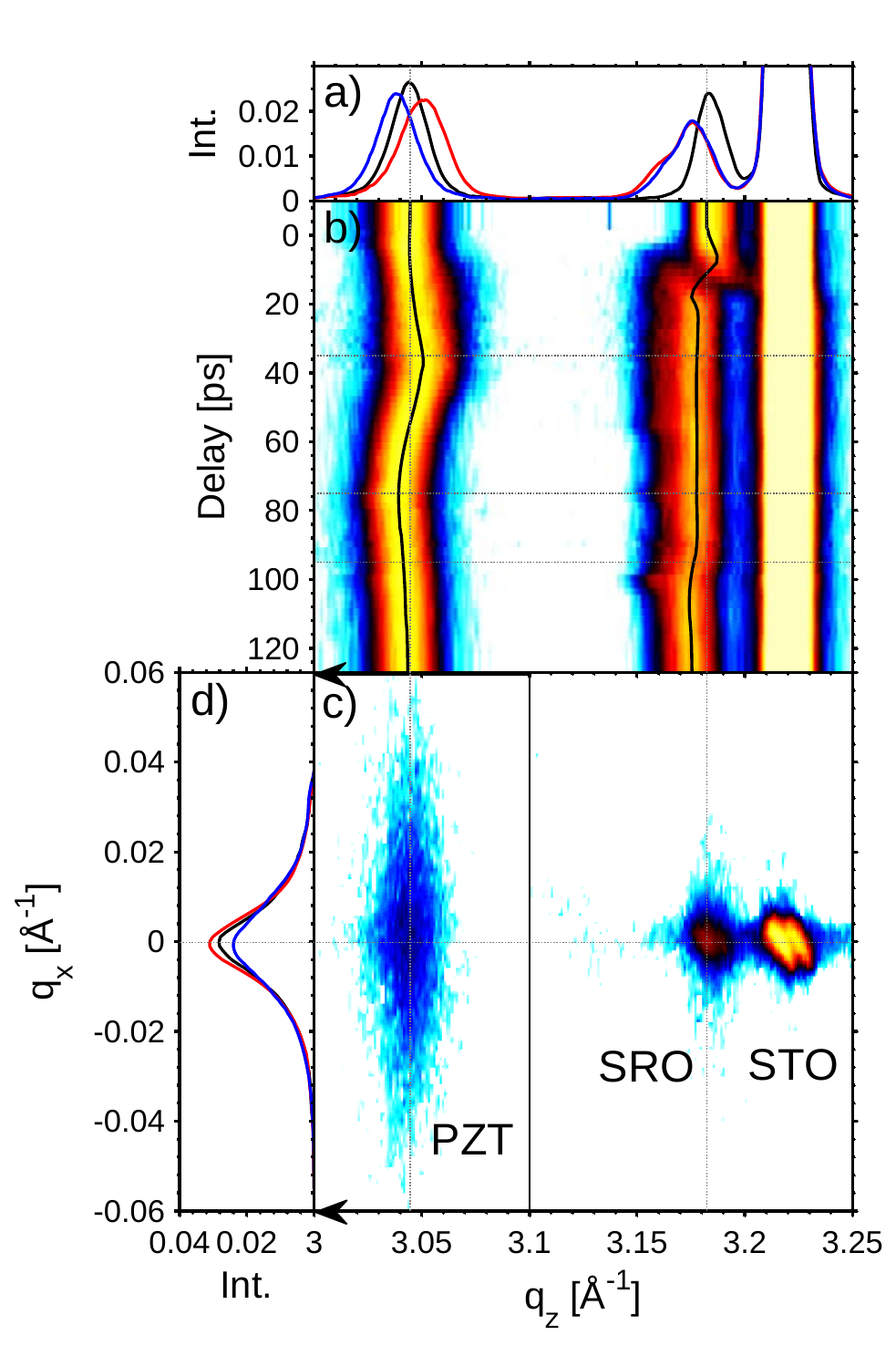}
\caption{Transient x-ray diffraction measurements and static reciprocal space map of the PZT-SRO sample.
a) $\theta-2\theta$ scans of the structure around the (0 0 2) Bragg reflections of PZT, SRO and STO before excitation (black) and at delays of maximum peak shift $s_z$ of PZT, at $t = 35$~ps (red) and $t = 75$~ps (blue).
b) Transient $\theta-2\theta$ scans for continuous variation of $-10 < t < 125$~ps. 
The horizontal dashed lines indicate the delays of the selected plots shown in panel a) and d).
The solid black lines indicate the center of the Bragg peaks.
c) The reciprocal space map of the PZT-SRO double layer sample before excitation features a rather broad PZT peak in $q_x$ direction. 
All peaks widths suffer from an additional broadening due to the convergence and energy bandwidth of the incident x-rays. 
d) Intensity of the PZT peak integrated over the $q_z$ dimension before excitation (black) and at delays of change of peak width $w_x$ of PZT, at $t = 35$~ps (red) and $t = 95$~ps (blue).}
\label{fig:rsm}
\end{figure}

As a typical structure, we grew a ferroelectric layer of PZT and a metallic SrRuO$_3$ (SRO) electrode layer onto an SrTiO$_3$ (STO) substrate by pulsed laser deposition (PLD).\cite{vrej2006b}
The transmission electron micrograph (TEM) image (Fig.~\ref{fig:temafm}~a) shows layer thicknesses of $d_\text{PZT} = 207$~nm and $d_\text{SRO} = 147$~nm, respectively.
The average lattice constants normal to the sample surface derived from static x-ray diffraction are $c_\text{PZT} = 4.130$~\AA\ and $c_\text{SRO} = 3.948$~\AA.
The TEM image (Fig.~\ref{fig:temafm}~a) features only few $a$-domains in the PZT layer, which are domains with a polarization vector pointing normal to the $c$-axis of the layer.\cite{vrej2006b}
Accordingly, out-of-plane polarized domains are called $c$-domains.
The small amount of $a$-domains is confirmed by the very weak scattering observed around 3.12~\AA$^{-1}$ in the static and transient rocking curves in Fig.~\ref{fig:rsm}~a-c).
We neglect these ferroelastic domains and switching between the $90^{\circ}$ polarizations states in the further discussion.
Due to stress relaxation in the mismatched epitaxial PZT film misfit dislocations (MD) at the SRO interface and threading dislocations (TD), expanding through the complete layer, are visible in Fig.~\ref{fig:temafm}~a).\cite{vrej2006b}
As a result lateral regions below 100~nm size are observable in PZT whereas the SRO layer is free of such inhomogeneities. 
Nevertheless, the AFM topography in Fig.~\ref{fig:temafm}~b) reveals that the mean roughness of the PZT surface is below 2~\AA.

In order to characterize the response of the PZT film to ultrashort stress pulses we excited the SRO electrode with near infrared (800~nm) femtosecond light pulses ($\tau=40$~fs) and monitored the induced lattice dynamics by UXRD experiments at a laser-driven plasma x-ray source (PXS)\cite{zamp2009a,schi2012a} in a pump-probe scheme.
The generated hard x-ray pulses ($E=8.05$~keV (Cu K$_\alpha$), $\tau=150$~fs) were collected by a Montel multilayer mirror and focused onto the sample with a convergence of $0.3^{\circ}$.
The diffracted photons were accumulated with a CMOS hybrid-pixel area detector in classical $\theta-2\theta$ geometry.
This allowed for detecting symmetrically and asymmetrically diffracted x-ray photons at the same time, avoiding time-consuming mesh scans in order to measure reciprocal space maps (RSM) around specific Bragg reflections.\cite{fews1997a,holy1999a,woit2005a,fews2004b}
Consequently, we acquired information both on in- and out-of-plane structure dynamics utilizing this time-resolved version of reciprocal-space mapping.
The temporal overlap of the optical pump and x-ray probe pulses was determined in an independent cross-correlation experiment\cite{boja2012a} and was set to the delay $0$~ps.


First we discuss the conventional x-ray diffraction from lattice planes parallel to the surface. 
Fig.~\ref{fig:rsm}~a) shows the $\theta-2\theta$ scans for three different time delays between optical pump and x-ray probe pulses. 
The black line represents the unexcited lattice and we can confirm the lattice constants of the three constituting materials from the Bragg angle.
The photoinduced dynamics are evident from the changes of the three material specific Bragg reflections, which are shown in Fig.~\ref{fig:rsm}~b).
By fitting each Bragg reflection for each material with a Gaussian we can extract the peak width $w_z(t)$ (FWHM) and peak center $c_z(t)$ for each delay $t$ in the $q_z$ dimension.
For the later analysis it is more convenient to introduce the peak shift $s_z(t) = c_z(t)-c_z(0)$.
The absorption of the pump pulse takes place exclusively in the SRO electrode layer, leading to a quasi-instantaneous temperature rise. 
The heat expansion of SRO by 0.35~\% is limited by the speed of sound in the material and proceeds within 24~ps evidenced by the shift to smaller Bragg angles.\cite{schi2012b} 
On the same timescale the substrate shows a tiny shoulder at the high-angle side according to the compression of STO adjacent to the expanding metal layer. 
Similarly, the PZT film is first compressed by the strain imposed from the expanding SRO, however it expands after the strain wave is reflected from the sample surface.\cite{Lee2005}
A detailed evaluation and discussion will be given below.

While this evaluation of UXRD signals is straightforward and the example shows the power of the method, we now discuss how to gain the information on the in-plane dynamics. 
The RSM before excitation is shown as a contour plot in Fig.~\ref{fig:rsm}~c).
In general, the size of the reciprocal lattice points in the RSM is inversely proportional to the length scale of coherently scattering regions of the crystal in the according in- and out-of-plane directions.
The additional broadening due to the instrument function of the x-ray diffraction setup, which is mainly given by the $0.3^{\circ}$ convergence and Cu K$_{\alpha}$ energy bandwidth of the incident x-rays, can be seen in the peak profile of the structurally perfect STO substrate in the RSM. 
Fig.~\ref{fig:rsm}~d) shows the diffraction signal integrated over the $q_z$ range of the PZT peak.
Similar to the convention above we define the width (FWHM) and shift for the $q_x$ dimension as $s_x(t)$ and $w_x(t)$. 
The large static value of $w_x^\text{PZT}$ is consistent with the average size of the lateral regions in the PZT layer of about 50~nm observed in the TEM image (Fig. \ref{fig:temafm}~a). 
In crystallography this broadening of $w_x$ can be described by the model of mosaicity\cite{alsn2001a}, assuming the crystal to consist of small mosaic blocks.
These blocks are homogeneous in themselves but the x-rays scattered from different blocks do not sum up coherently. 
The in-plane size of the blocks 
defines the lateral correlation length which is inversely proportional to the broadening of the RSM in $q_x$.
Tilting of the blocks can give rise to an additional broadening. These two effects may be distinguished by measuring a RSM around an asymmetrical Bragg reflection.\cite{fews1997a}
Fig.~\ref{fig:rsm}~d) shows that $w_x$ increases considerably for snapshots recorded after the reflection of the strain wave at the sample surface.

\begin{figure}
\includegraphics[width=0.9\columnwidth]{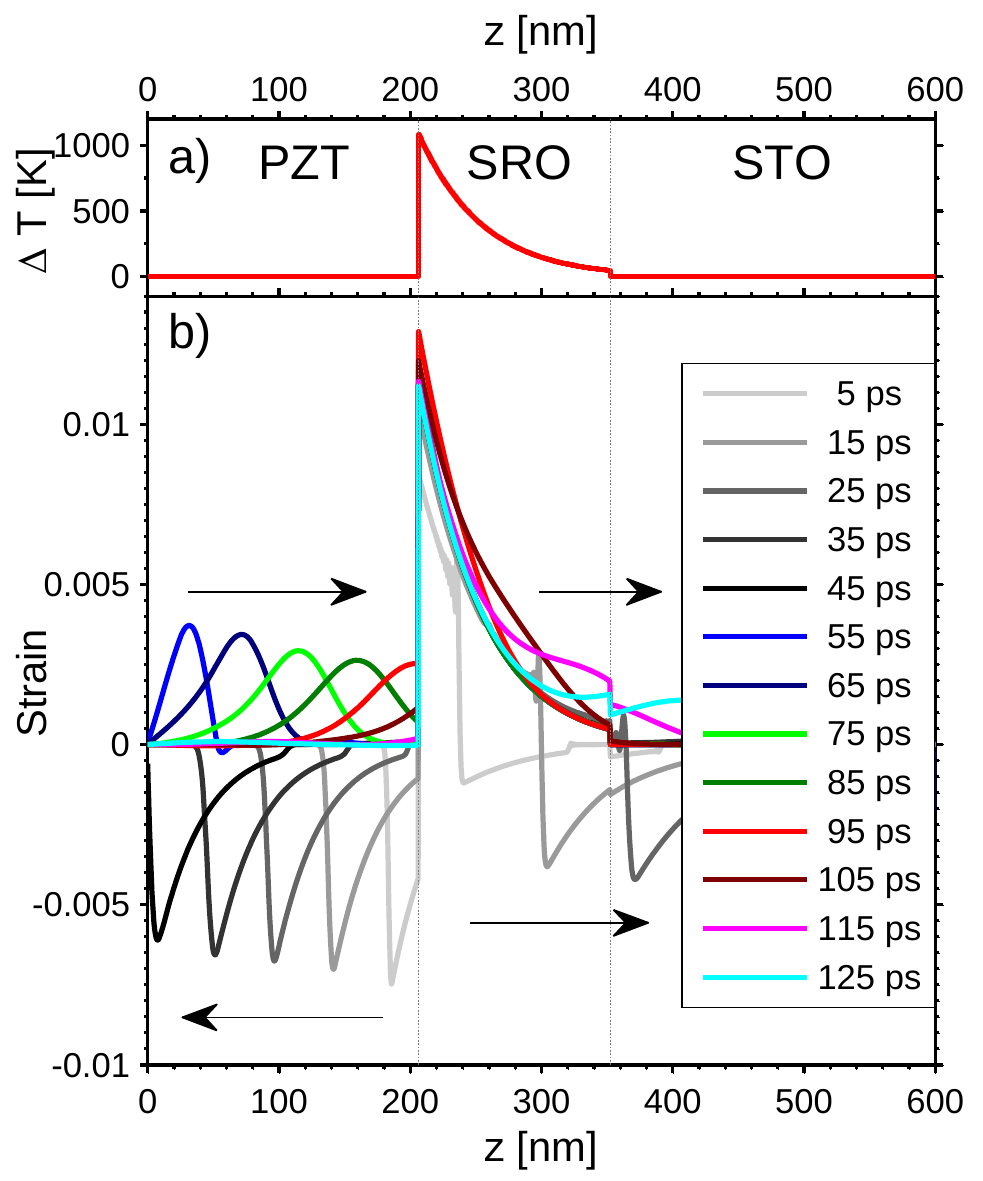}
\caption{Simulation results for the temperature gradient and lattice dynamics after optical excitation.
a) The optical pump laser causes a quasi instantaneous temperature increase only in the metallic SRO layer at delay zero.
b) The excited lattice dynamics are calculated by a 1D linear chain model. 
The strain profiles show a coherent soundwave with a sharp leading edge travelling from the PZT-SRO interface to the PZT-air interface.
It is converted to an expansion wave with much stronger damping, smoothing and broadening the profile.}
\label{fig:excphonons}
\end{figure}

In order to discuss our experimental results we apply a 1D model of the sample structure to simulate the lattice dynamics by a linear chain model of masses and springs.\cite{herz2012b} 
These simulations are well established to predict the out-of-plane dynamics but do not consider the in-plane dynamics directly.
Therefore, we employ the out-of-plane phonon damping as an adjustable parameter to couple energy to in-plane motion.
First, we calculate the temperature rise in the SRO after optical excitation (Fig. \ref{fig:excphonons}~a) from the laser fluence, absorption depth and heat capacity of this metal.
The quasi-instantaneous thermal stress excites coherent acoustic phonons (strain waves) which are launched from the interfaces to the PZT layer and STO substrate. 
Calculated strain profiles for different delays are depicted in Fig. \ref{fig:excphonons}~b).
Since heat diffusion from the SRO into the PZT layer can be neglected on this short time scales\cite{shay2011a} ($\approx$ 100~ps) the PZT lattice dynamics are exclusively determined by the compression wave travelling from the SRO-PZT interface to the PZT-air interface. 
Here the strain wave is reflected and converted into an expansion wave travelling back to the SRO layer and further into the STO substrate.
Due to the good acoustic impedance matching of the three materials we neglect reflections at layer interfaces.

\begin{figure*}
\includegraphics[width=0.9\textwidth]{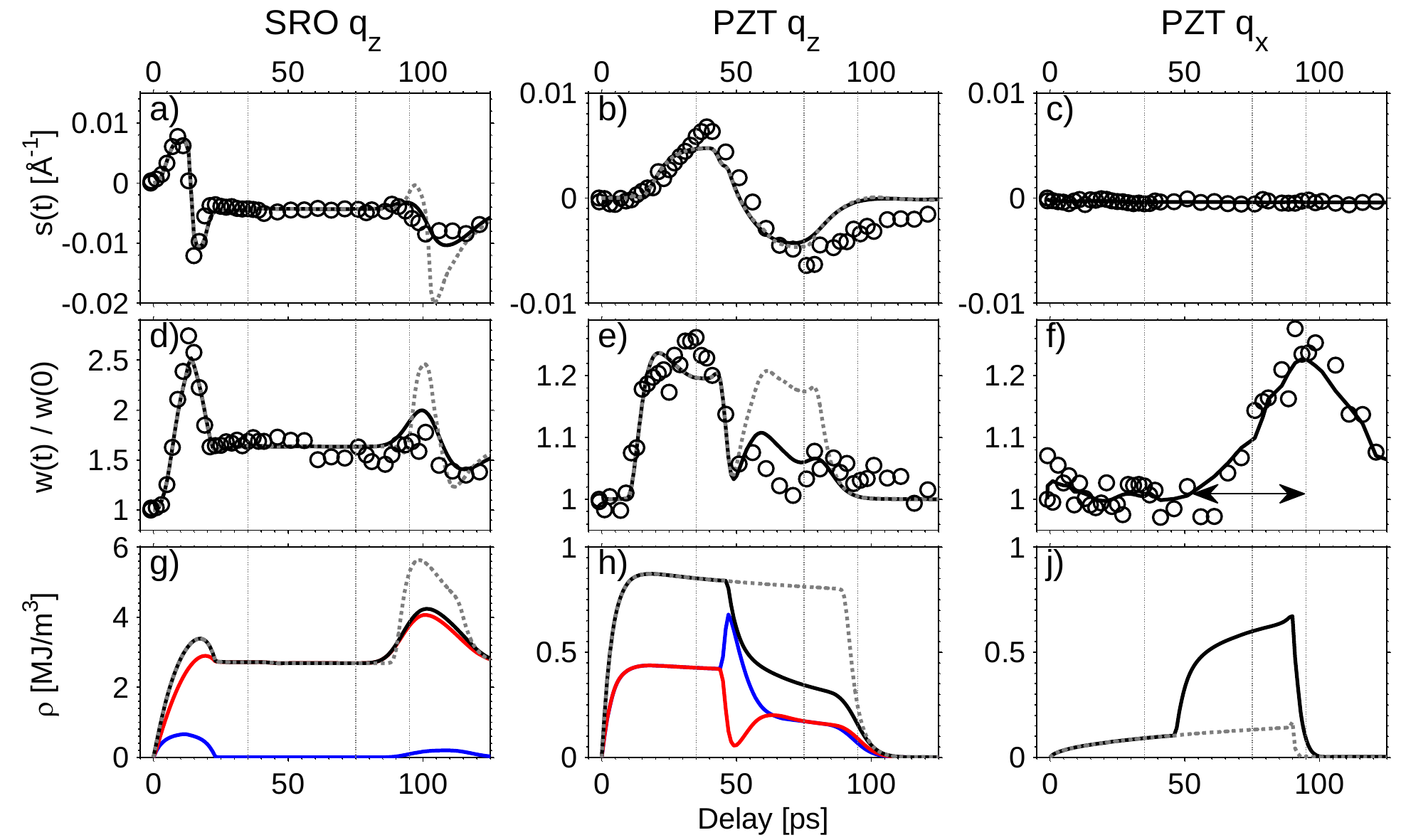}
\caption{a-f) Comparison of experimental data (circles) with the simulation results (solid black lines).
In Panel c) and f) the solid lines are only guides to the eye.
All grey dashed lines are the analogue to the black lines assuming constant damping in PZT.
Panels g) and h) show the kinetic (blue), potential (red) and total (black) energy density in the according layer.
j) Damped energy density of out-of-pane motion in PZT.}
\label{fig:shiftwidth}
\end{figure*}

From the simulated spatio-temporal strain map, the resulting transient changes of the x-ray diffraction profile in $q_z$ dimension for the PZT and SRO layers are calculated by dynamical x-ray diffraction theory.\cite{herz2012a} 
Taking the elastic constants of each material, we can use the damping of coherent phonons by impurities and coupling to in-plane motion as adjustable parameters. 
Fig.~\ref{fig:shiftwidth}~a,b,d,e) show the excellent agreement of the simulated x-ray diffraction data with the measured values for $s_z$ and $w_z$.

The peak shift $s_z$ is a measure of the average $c$-axis lattice parameter of the PZT and SRO layers, which was qualitatively discussed above. 
The change of $w_z$ essentially reflects the inhomogeneous strain, which in SRO is given by the short absorption length of the optical pump light leading to a stress exponentially decaying with $z$. 
Initially SRO only expands near the PZT interface.
At 12~ps after excitation, the expansion wave has propagated through half the SRO layer, which leads to a maximum $w_z^\text{SRO}$ (Fig. \ref{fig:shiftwidth}~d), which in fact reflects a splitting of the SRO Bragg peak.\cite{herz2012b,schi2012b}
Due to the peak splitting the Gaussian fit indicates a compression of the SRO layer for $s_z^\text{SRO}$, as long as only a small fraction of SRO is expanded.
Similarly to SRO, $w_z^\text{PZT}$ rises and the peak shifts to larger angles as the compression travels through the PZT layer (Fig. \ref{fig:shiftwidth}~b,e). 
Later $s_z^\text{PZT}$ becomes negative when the soundwave is reflected at the surface ($t > 50$~ps) and propagates back to the substrate.
$w_z^\text{PZT}$ is much smaller for $t > 50$~ps although the tensile strain during expansion, $s_z^\text{PZT}$, has nearly the same magnitude as the preceding compression.
This implies that the strain pulse in PZT broadens in space rendering the layer less inhomogeneously strained when the layer expands.

In order to fit these four transient data sets for PZT and SRO (Fig. \ref{fig:shiftwidth}~a,b,d,e) simultaneously in our simulations we have introduced phonon damping in PZT as the only free parameter.
The best fit for the four data sets was achieved when the damping factor in PZT is chosen 50 times larger for expansive out-of-plane strain compared to compressive strain of the same magnitude.
The according results are plotted as solid black lines in Fig. \ref{fig:shiftwidth}~a,b,d,e).
The grey dashed lines represent the results for a constant damping factor in PZT for expansion and compression.
We can exclude pure surface scattering as the reason for the asymmetric damping behavior, since this would lead to a much smaller amplitude of $s_z^\text{PZT}$ in order to achieve the same decrease of $w_z^\text{PZT}$ for the expansive strain.
The increase of the damping in PZT is visualized in Fig.~\ref{fig:excphonons}~b) where the coloured lines ($t>50$~ps) show a smooth out-of-plane expansion, whereas the greyish lines ($t<50$~ps) show the inhomogeneous compression before the reflection at the surface changes the sign of the strain wave.

From our lattice dynamics simulations we can also determine the kinetic (blue), potential (red) and total (black) energy density of the out-of-plane coherent phonons in each layer (Fig. \ref{fig:shiftwidth}~g-h).
We introduced the phonon damping in PZT to couple energy to lateral phonons.
This energy is essentially the difference of the total energy of out-of-plane coherent phonons in PZT with and without damping.
The result is plotted in Fig. \ref{fig:shiftwidth}~j) where the grey dashed line corresponds again to the case of constant damping.
The increase of the lateral energy density in PZT (Fig. \ref{fig:shiftwidth}~j) goes along with a considerable increase of $w_x^\text{PZT}$ which reflects a change of the inhomogeneity in plane, probably because the lateral blocks develop an inhomogeneous in-plane strain that is dynamically coupled to the out-of-plane motion according to the Poisson ratio (Fig. \ref{fig:shiftwidth}~f). 
We do not observe the converse effect during the compression of the PZT layer ($t<50$~ps).
The horizontal arrow in Fig. \ref{fig:shiftwidth}~f) indicates the timescale of the build-up of the lateral strain of approx. 50~ps.
We can link this timescale to a lateral lengthscale of approx. 200~nm via the sound velocity of PZT of 4.6~nm/ps.
This timescale agrees well with the in-plane block size observed in the TEM image in Fig. \ref{fig:temafm}~a) and $w_x^\text{PZT}$ of the static RSM in Fig. \ref{fig:rsm}~c).

The analysis of the measured signal alone already suggests the following interpretation: 
The expansion of SRO sends a compression wave into PZT. 
The in-plane mosaicity/nanoinhomogeneity is unchanged during this period.
When the strain changes sign upon reflection at the surface PZT expands and according to Poisson ratio, the mosaic blocks must now laterally contract. 
The in-plane inhomogeneity is increased as millions of in-plane contraction waves start at all the lateral dislocations.
We conclude that only out-of-plane \textit{expansion} of PZT couples energy to in-plane dynamics, and that this effect is essentially suppressed for out-of-plane compression, since this would have to expand the mosaic blocks, which is sterically forbidden by the adjacent blocks.
This compares favourably with our simulations of the out-of-plane lattice dynamics, requiring an increased damping for the expansion wave in the PZT, which can be understood as an increase of the coupling between in- and out-of plane lattice motions. 

In conclusion we have demonstrated the first measurement of the lattice dynamics in a structurally imperfect thin film by ultrafast reciprocal space mapping (URSM).
We not only extract the changes of the lattice constants, i.e. the expansion and compression of materials perpendicular to the surface.
In addition, we quantify the coupled response in plane, which turns out to be significantly enhanced for out-of-plan expansion, as it provides the in-plane contraction necessary for the atoms to start moving.
URSM will be an important method for understanding the ultrafast response of oxide crystals with their natural tendency to form nanoscale inhomogeneities.

We thank B. Birajdar (Max-Planck-Institut f\"ur Mikrostrukturphysik, Halle, Germany) for performing the TEM imaging.
This work was supported by the Deutsche Forschungsgemeinschaft via grant No. BA 2281/3-1 and by the German Bundesministerium f\"ur Bildung und Forschung via grant No. 03WKP03A and 05K10IP1.


%

\end{document}